\documentclass[smallextended, final]{svjour3}
\usepackage[margin=1in]{geometry}
\usepackage{comment}

\date{\large This work was submitted to \href{https://www.springer.com/journal/40192}{Integrating Materials and Manufacturing Innovation} on July 7, 2020 and is currently under peer review.}
\author{Ryan Cohn, Elizabeth Holm}

\usepackage[sorting=none, bibencoding=utf8, backend=bibtex, style=numeric-comp]{biblatex}
\bibliography{refs_update}

\usepackage{amsmath}

\usepackage{xcolor}
\usepackage{hyperref}

\usepackage{graphicx}
\graphicspath{Figures}

\title{Unsupervised machine learning via transfer learning and k-means clustering to classify materials image data}
\author{Ryan Cohn \and Elizabeth Holm}
\institute{Materials Science and Engineering 
Carnegie Mellon University 
5000 Forbes Ave., Pittsburgh, PA 15213, USA}

\begin{document}

\maketitle

\begin{abstract}
    Unsupervised machine learning offers significant opportunities for extracting knowledge from unlabeled data sets and for achieving maximum machine learning performance. This paper demonstrates how to construct, use, and evaluate a high performance unsupervised machine learning system for classifying images in a popular microstructural dataset. The Northeastern University Steel Surface Defects Database includes micrographs of six different defects observed on hot-rolled steel in a format that is convenient for training and evaluating models for image classification. We use the VGG16 convolutional neural network pre-trained on the ImageNet dataset of natural images to extract feature representations for each micrograph. After applying principal component analysis to extract signal from the feature descriptors, we use k-means clustering to classify the images without needing labeled training data. The approach achieves \(99.4\% \pm 0.16\%\) accuracy, and the resulting model can be used to classify new images without retraining This approach demonstrates an improvement in both performance and utility compared to a previous study. A sensitivity analysis is conducted to better understand the influence of each step on the classification performance. The results provide insight toward applying unsupervised machine learning techniques to problems of interest in materials science.
    \\
    \\
    \noindent All code used in this study is available on GitHub in the following repository: 
    
    \noindent \href{https://github.com/rccohn/NEU-Cluster}{https://github.com/rccohn/NEU-Cluster}
    \\
    \\

    \keywords{computer vision \and transfer learning \and image classification \and convolutional neural network \and unsupervised machine learning}
\end{abstract}

\section{Introduction}


While applications of machine learning to materials science problems currently focus on supervised machine learning \cite{Dimiduk2018}, there are significant opportunities for unsupervised machine learning, both for extracting knowledge from unlabeled data sets and for achieving maximum machine learning performance. However, because they are less prevalent in the background literature, the best practices – and pitfalls – of using unsupervised methods are often overlooked. In this paper, we demonstrate how to construct, use, and evaluate a high performance unsupervised machine learning approach to a standard materials computer vision problem. We explain each step in the process, including a sensitivity analysis of user-selected parameters, and we publish our code as supplementary information. The code is available on GitHub in the following repository: \href{https://github.com/rccohn/NEU-Cluster}{https://github.com/rccohn/NEU-Cluster}.

There is growing interest in employing computer vision for the quantitative analysis of microscopy images in materials science \cite{Holm2020,DeCost-chapter}. Potential applications of computer vision are extensive in both academic research and industrial materials processing. Recent studies have shown that computer vision approaches can be used for tasks including interpreting diffraction patterns \cite{Ram2017, Ziletti2018}, process control and powder characterization for additive manufacturing \cite{Scime2018, TanPhuc2019, DeCost2017a}, automatic detection of features of interest in micrographs \cite{Kusche2019, Campbell2018, Jiang2017, Chen2018, DeCost2017c}, and image segmentation and quantification\cite{DeCost-segment}, among others. With this in mind, we select a computer vision problem as the exemplar for a high performance unsupervised machine learning system.

Image classification, the process of assigning a discrete label to an image to describe its contents, is a fundamental task in computer vision. The ImageNet Large Scale Visual Recognition Challenge (ILSVRC) is one of the biggest standardized efforts to evaluate computer vision approaches to natural image classification (i.e. photos of soccer balls, bassoons, hummingbirds, etc.) \cite{Russakovsky2015}. Recent developments in deep learning with convolutional neural networks (CNNs) \cite{Simonyan2014a, He2016}  have been shown to approach and even surpass human performance on classifying images in the ImageNet database \cite{He2015}. There are numerous CNN architectures, but they all include layers of filters that encode quantitative descriptions of the visual contents of the image. 

Training a CNN from its original randomized weights generally requires a very large training set of labeled images. For example, the ImageNet 2014 dataset contained over 1,200,000 labeled images in the training set. For the materials scientist, gathering data at this scale is prohibihitively expesive and time consuming. Interestingly, however, the intermediate layers of a trained CNN can be used to extract meaningful features for images that do not match the class descriptions in the original training set.  This process, called transfer learning, allows the materials scientist to use CNNs trained on large databases of natural images, like ImageNet, for the task of quantitative microstructural image analysis \cite{Kitahara2018, Ling2017}, without retraining on any additional images.

After extracting visual features for each image, the typical approach to image classification uses supervised learning. In this process, a classifier is shown labeled examples in the training set in order to learn the decision boundaries between images of each class.  After training the classifier can predict the labels of new images. In contrast, clustering, or unsupervised learning, is the process of finding natural patterns in the data to determine class labels without the use of labeled training data. Clustering is useful when labeling data is costly or when class labels are difficult for a human to define. An additional advantage is the ability to learn from the entire dataset, without the requirement to hold out data for validation and testing. Because it finds natural patterns in the data, the labels determined by clustering are not guaranteed to agree with the class labels assigned by humans. However, when the feature descriptors capture the relevant visual signals in each image, clustering can be used to classify data with very high performance.

\begin{figure}[bt]
    \centering
    \includegraphics[scale=0.5]{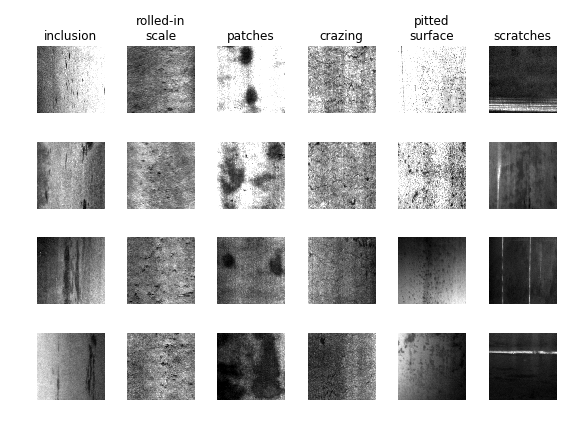}
    \caption{Sample images from Northeastern University Steel Surface Defects Database. Each column contains images from a different defect class in the dataset.}
    \label{fig:SDD}
\end{figure}

The Northeastern University Steel Surface Defects Database \cite{Song2013} (NEU-SSDD) is an open-source database containing labeled images of steel defects.  It is quickly becoming a standard for evaluating the performance of image classification and object detection algorithms applied to image data in materials science with many citations in the literature \cite{Kitahara2018, Song2013, Yi2017zz, Gao2020zz, Liu2017zz, Luo2019zz, Ren2018zz, Tao2018zz, Zhou2017zz, Xiao2017zz}. The NEU-SSDD consists of 1800 images of surface defects observed on samples of hot-rolled steel. The images are evenly divided into six classes. The defect classes are inclusions, rolled-in scale, patches, crazing, pitted surfaces, and scratches. All images are the same size and format making them convenient for analysis. Example images from the database are shown in Figure \ref{fig:SDD}. 

Because the NEU-SSDD images are all labeled with a defect type, they are well-suited to supervised machine learning, and indeed supervised ML methods have achieved excellent classification accuracy as high as 99\% \cite{Zhou2017zz}. However, the NEU-SSDD also offers a testbed to compare supervised and unsupervised methods for image classification. In 2018, Kitahara used transfer learning with the VGG16 network coupled with unsupervised k-means clustering to classify images in the NEU-SSDD with \(98.3\% \pm 1.2\%\) accuracy \cite{Kitahara2018}. In their approach, features were projected onto t-distributed stochastic neighborhood embedding (t-SNE) maps before clustering. Since t-SNE maps cannot incorporate new points, their method cannot be extended to classify additional images. In this study, we adopt a similar strategy for classifying the NEU-SSDD dataset. However, by clustering whitened principal components instead of t-SNE embeddings, and by increasing the number of iterations for which k-means is run, we classify the data with higher performance and better repeatability between trials. Additionally, our method can be used to classify new data, which is important for applications in high throughput experiments or quality control settings. Since seemingly small changes to the approach result in big changes in utility and classification performance, we conduct a sensitivity analysis to demonstrate the impact of each step on the results.

\section{Methodology}

The analysis is split into 3 parts: 
\begin{enumerate}
    \item Pre-processing: Prepare the data to be read by the CNN. 
    \item Feature extraction (encoding): Use the CNN to generate a numerical representation of each image. 
    \item Clustering (decoding): Assign a label to the image, grouping images with similar features together.   
\end{enumerate}

Python was used to conduct the study. Scikit-image \cite{VanderWalt2014} was used for preprocessing image data. The VGG16 neural network \cite{Simonyan2014a} pretrained on the ImageNet dataset \cite{Russakovsky2015}, accessed through Keras \cite{Chollet2015}, was used to extract features. Scikit-lean \cite{Pedregosa2011} was used to further process the data and perform the clustering. These tools are described in detail in the following sections, and Python code is included in the supplementary online materials.

\subsection{Pre-processing}
\label{ss:pp}

\begin{figure}[tb]
    \centering
    \includegraphics{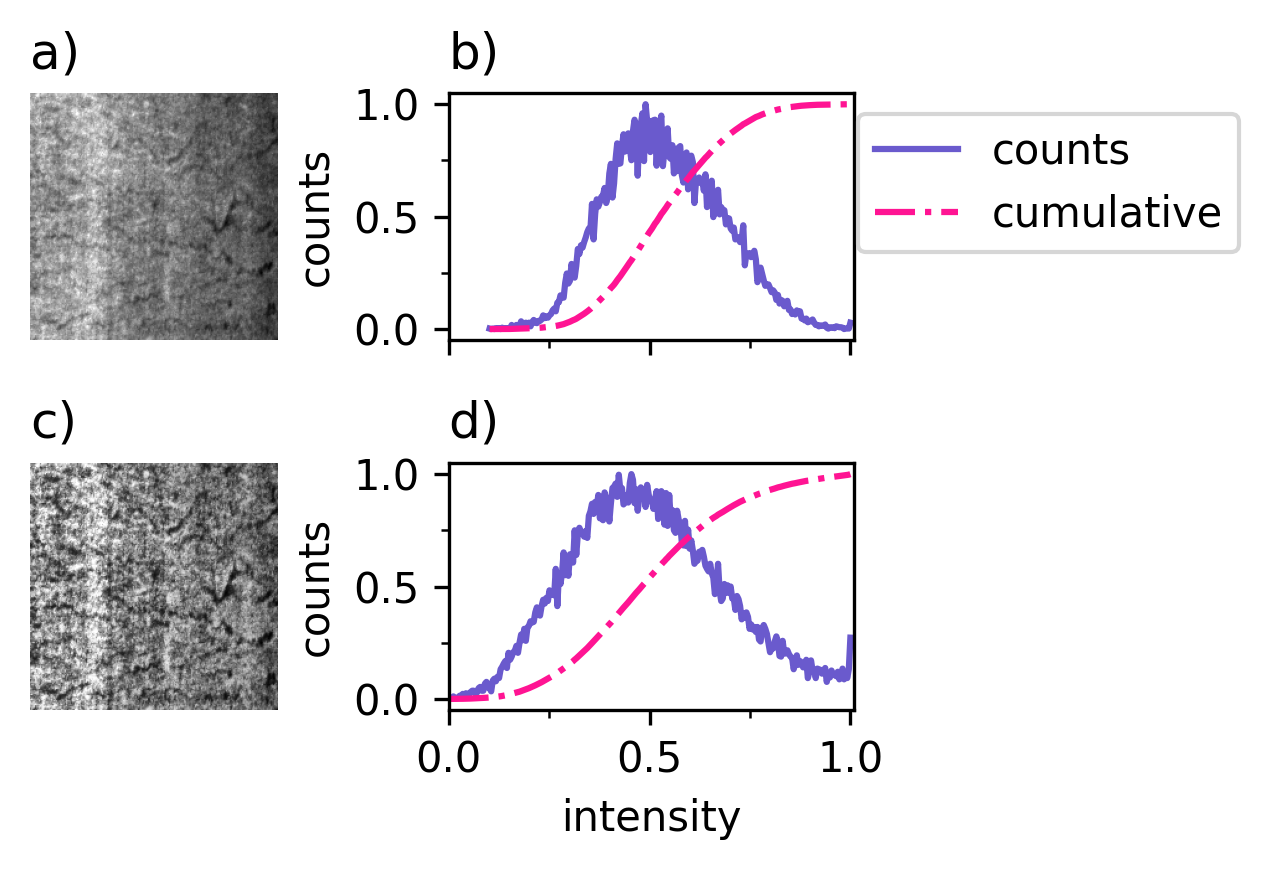}
    \caption{CR\_10 from the NEUS-SDD dataset. a) Original image. b) Intensity histogram of original image. c) Image after CLAHE applied.  d) Intensity histogram of image after CLAHE applied.}
    \label{fig:HistEQSample}
\end{figure}
The first step to image analysis is loading an image file into Python. This can be done in scikit-image using the command \textbf{skimage.io.imread()}. After loading an image, contrast-limited adaptive histogram equalization (CLAHE) \cite{Pizer} is applied. Histogram equalization normalizes the intensity (brightness) distribution of each image. This can make the analysis more robust to differences in lighting conditions between images. Additionally, this technique can increase the contrast of different regions in the image, leading to stronger responses from internal filters in the CNN. (Note, however, that histogram equalization is not always helpful, since it may eliminate a relevant brightness signal or over-emphasize certain image features.) In scikit-image, CLAHE can be applied to an image using \textbf{skimage.exposure.equalize\_adapthist()}. An example of CLAHE applied to an image from the NEU-SSDD is shown in Figure \ref{fig:HistEQSample}. After the histogram equalization is applied, the intensity distribution is much wider, and crazing in the image is much more noticeable.


The next step for pre-processing the image is resizing. Because the weights of a neural network are trained using training images of a fixed size, the network is only able to process images with the same dimensions. Images from the NEU-SSDD are scaled up from 200x200px to 224x224px for VGG16. Interpolation is applied to images before resizing to allow for rescaling by a non-integer scaling factor. To resize images with scikit-image, the command \textbf{skimage.transform.resize()} was used. After pre-processing was complete the images were saved to disk using the \textbf{skimage.io.imsave()} command.

\subsection{Feature Extraction}
The VGG16 network \cite{Simonyan2014a}, developed by the Oxford Visual Geometry Group, is a popular CNN for computer vision tasks because of its high performance and relative simplicity. The VGG16 network achieved the top score in the ILVRC 2014 challenge, and is still widely used for computer vision tasks today. Thus, we use the VGG16 network to assist with the task of classifying images of steel defects in this study. Images were read from the disk using the function \textbf{keras.preprocessing.image.load\_img()}. Images were then formatted using the function \textbf{keras.applications.vgg16.preprocess\_input()}. Note that unlike pre-processing described in Section \ref{ss:pp} these steps ensure the images are formatted correctly for use with the VGG16 model and do not change the properties of the images themselves.

After reading the images, the VGG16 network pre-trained on the ImageNet database was used to extract numerical feature descriptors of each image. The pre-trained network can be accessed through \textbf{keras.applications.vgg16.VGG16()} with the keyword argument \textbf{weights=imagenet}.The images are fed into the input layer of the neural network. The outputs of one or more intermediate layers in the network can be used as the feature representation of the image. Due to the black box nature of neural networks, the task of choosing which layer or combination of layer outputs to use is a matter of trial and error, and is an area of active research \cite{Ling2017}. In many cases, using the output of a single convolution layer or fully-connected layer are sufficient for the task of image classification. In this study, the FC1 layer of VGG16 was used as it was empirically found to have good performance \cite{Kitahara2018}.

\begin{figure}[tb]
    \centering
    \includegraphics[scale=1]{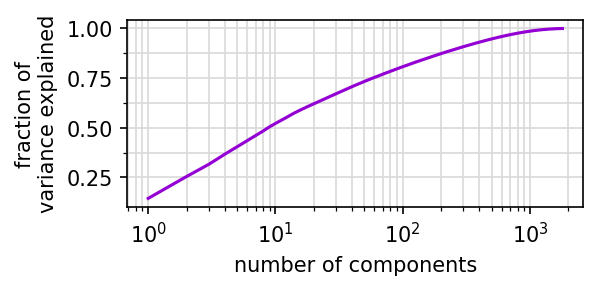}
    \caption{Cumulative fraction of variance explained vs number of PCA components for FC1 features of NEU-SSDD dataset.}
    \label{fig:NumComponents}
\end{figure}

The outputs of intermediate layers in VGG16 each contain between 4,096 and 3,211,264 elements. These outputs include a significant amount of noise and zero-elements resulting from filters that did not activate. Principal component analysis (PCA) \cite{Jollife2016} is applied to reduce the dimensionality of the data, simultaneously increasing classification performance while decreasing computational cost. First, the data are centered so the mean along each dimension is zero. Next, the covariance matrix is computed from the centered data. Finally, singular value decomposition (SVD) is performed to determine the eigenvectors and eigenvalues of the covariance matrix. The eigenvectors are the principal components which the data can be projected onto. The eigenvalue corresponding to each eigenvector is proportional to the amount of variance explained by that component. 


To reduce the dimensionality of the data, the data are projected onto the n principle components that explain the highest fraction of variance. Selecting n is subjective, but reasonable values can be determined from the amount of variance explained by each component. The fraction of variance explained versus number of PCA components used for FC1 features on the NEU-SSDD dataset is shown in Figure \ref{fig:NumComponents}. Using 100 components explains about 80\% of the variance in the data. The remaining 1700 components only capture 20\% of the variance and are likely not needed for classification. In fact, using too many components can actually decrease classification performance by introducing noise. The number of components to use can also be selected by their impact on the `natural' clustering of the data. This is described more in Section \ref{ss:methods:clustering}.

After selecting the number of components, whitening is an optional post-processing step for PCA. Whitening scales the final subset of components by the variance of each component to achieve unit variance across all components. This may improve the classification performance. However, if too many components are chosen, whitening the data may decrease performance due to increasing the weight of noisy components. The effect of whitening on classification performance is discussed in further detail in Section \ref{sec:results-pca}. In Python, Scikit-learn provides a convenient implementation for PCA, including whitening, with the  \textbf{sklearn.decomposition.PCA()} object.

\subsection{Clustering}
\label{ss:methods:clustering}
The last step in the analysis is clustering, in which the class label for each image is assigned. K-means clustering \cite{Lloyd1982} is an unsupervised machine learning method that is one of the most popular clustering algorithms in use today \cite{Bock2007, Jain2010}. The goal of k-means is to group nearby points in feature space.  The k-means clustering algorithm works in the following way:
\begin{enumerate}
    \item Choose k centroids.
    \item \label{step:: associate} Associate each point with the centroid that is closest to it in feature space.
    \item \label{step:: update}  Update the position of each centroid to be the mean position of all of the points associated with it.
    \item Repeat steps \ref{step:: associate} and \ref{step:: update} until the centroids do not change position or until a maximum number of iterations is reached.
\end{enumerate}

\begin{figure}
    \centering
    \includegraphics{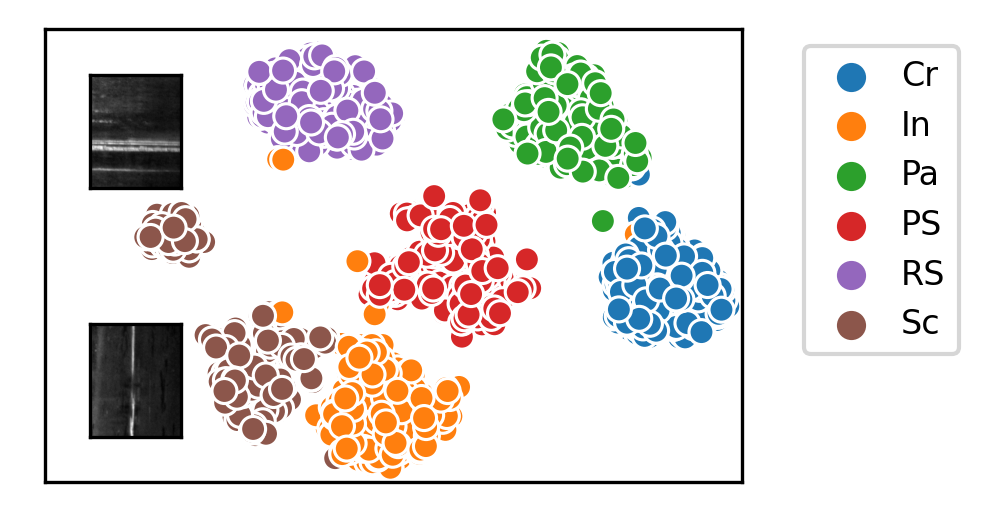}
    \caption{t-SNE map for unwhitened PCA components. Points in the map are colored by their ground truth labels and show good clustering.  Typical images for the two different clusters corresponding to scratches are overlaid next to each cluster and show the separation of vertical and horizontal scratches. }
    \label{fig:tsne_scratch}
\end{figure}

This approach minimizes inertia, which is the sum of squared euclidean distances from each point to its associated cluster center. Note that this achieves a local minimum that is dependent on the initial position of each centroid. There are different approaches for selecting the starting centroids. One method, called k-means++ \cite{Arthur}, has been shown to help k-means achieve good clustering performance and computational efficiency. In this technique, the first cluster center is chosen with uniform probability from the data. The remaining cluster centers are chosen from the data with probability proportional to the distance to the nearest cluster center. Thus, the initial cluster centers are close to data points and are spread out from each other, which is consistent with the expected patterns in the data when using k-means clustering.

Once the initial centroids are selected, k-means is deterministic. However, selecting the initial cluster centers is a stochastic process. Unless the data naturally cluster very well and do not contain noise or outliers, running k-means with different initializations will give different results. Therefore, k-means is run several times with different initial centroids. The cluster centers with the lowest inertia are used as the final clusters. Note that inertia is used instead of accuracy to select the clusters. This prevents overfitting the model and allows for clusters to be determined without the use of labeled training data. This is discussed in more detail in Section \ref{ss:cluster}.

K-means requires an input value for K, the number of clusters. There are several techniques for approximating a reasonable value for K \cite{Charrad2014, Pelleg2000}. The number of clusters can also be determined empirically from visualizing the data. t-Distributed Stochastic Neighbor Embedding (t-SNE) is a popular method for visualizing high-dimensional data. This technique is introduced in \cite{Maaten2008} and summarized in \cite{Kitahara2018}. t-SNE maps for the NEU-SSDD data are shown throughout this paper. In this technique, a nonlinear transformation is used to project data to 2 or 3 dimensions so it can be analyzed visually. The distances between points that are close to each other on the resulting t-SNE maps are more likely to be representative of the actual distances in the original feature space. The distances between points that are far away from each other on t-SNE maps are not likely to be representative of the actual distances in feature space. In other words, points that cluster together in feature space are likely to cluster together on t-SNE maps. The spacing between different clusters may not be representative of the actual inter-cluster spacing, but different clusters  are still distinct. 

A sample t-SNE map for the NEU-SSDD is shown in Figure \ref{fig:tsne_scratch}. The map was computed using PCA components without whitening, which shows more distinct clusters than the map computed from whitened components.
The map reveals the natural clustering of the data. In this figure, the numerical values on the axes are not shown because they are arbitrarily determined during the projection. Since there are 6 defect classes,  at least 6 clusters must be used to classify the data. However, on the t-SNE map, points corresponding to scratches appear to be separated into two distinct clusters. Representative images for each of the clusters corresponding to scratches are overlaid on the t-SNE map next to their respective clusters. From these images it is clear that these clusters distinguish horizontally oriented scratches from vertically oriented scratches. Neural networks are commonly sensitive to rotation, and reducing this phenomena is an area of active research \cite{Cheng2016}. Thus, for this dataset, adding a 7\textsuperscript{th} class, distinguishing between between horizontal and vertical scratches, is necessary to maximize the classification performance. 

In \cite{Kitahara2018}, clustering is performed directly on t-SNE mappings of the NEU-SSDD data. This approach has the advantage of being interpretable as clusters can be visualized directly. However, this method will not generalize to new data as t-SNE maps must be recomputed every time new data is added. In this paper it is shown that clustering directly on whitened PCA components can achieve comparable performance to clustering the t-SNE maps and can scale to larger datasets and generalize to new data more effectively.

Scikit-learn provides a convenient implementation of k-means with the object \\ \textbf{sklearn.cluster.KMeans().} This implementation contains the k-means++ algorithm for centroid initialization with the \textbf{init=`k-means++'} keyword argument.

\begin{figure}[bt]
    \centering
    \includegraphics{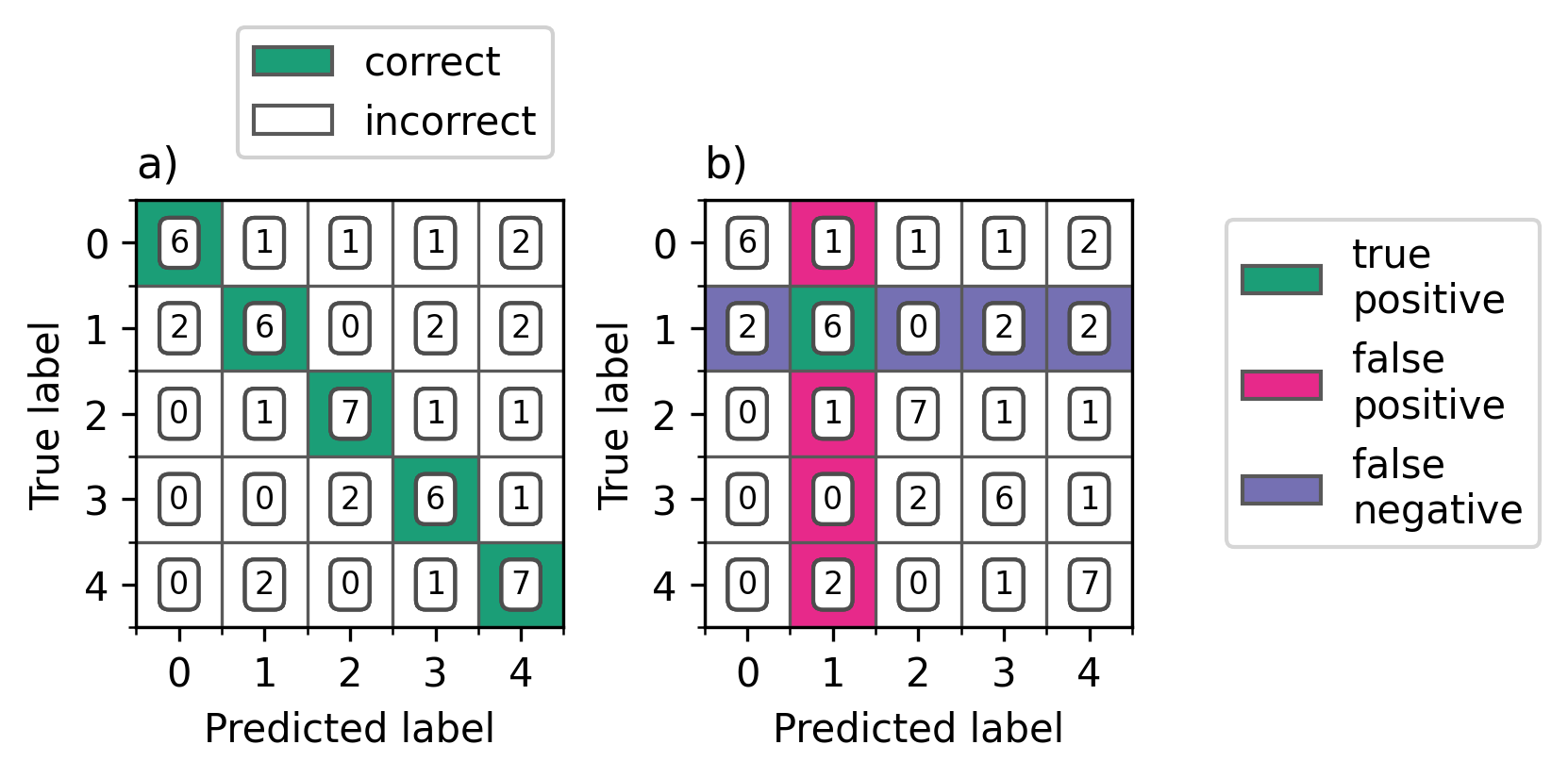}
    \caption{Sample confusion matrix. a) Correct predictions are highlighted. Incorrect predictions are shown as off-diagonal elements. b) Results for class 1 are highlighted. True positive predictions for class 1 are in element (1,1). False positives and false negatives for class 1 are other elements in the \(\text{1}^{\text{st}}\) column and \(\text{1}^{\text{st}} \) row, respectively. }
    \label{fig:CM}
\end{figure}
\subsection{Model Evaluation}
When ground truth class labels are available, clustering algorithms can be evaluated on the basis of accuracy, precision, and recall. Accuracy is the ratio of predictions made by the model that are correct. Accuracy quantifies the performance of all predictions into a single value, but can be misleading for unbalanced datasets. Precision and recall contain more information about the model performance but must be calculated for each ground-truth class in the dataset.  The precision of predictions for class \(i\) describes the likelihood that a prediction of class \(i\) made by the model is correct. The recall for class \(i\) describes the likelihood that a datapoint with ground-truth class \(i\) will be correctly labeled by the model. These values are  determined from true positives, false positives, and false negatives, which are defined in Table \ref{tab:tpfpfn}. Precision and recall are defined as
\begin{equation}
    \label{eqn:precision}
    P_i = \frac{TP_i}{TP_i+FP_i}
\end{equation}
\begin{equation}
\label{eqn:recall}
    R_i = \frac{TP_i}{TP_i+FN_i}
\end{equation}
where \(P_i\) and \(R_i\) are the precision and recall for ground-truth class \(i\), respectively. \(TP_i\) and \(FP_i\) are the number of true positive and false positive predictions for class \(i\), respectively.

\begin{table}[tb]
    \centering
    \begin{tabular}{l c c}
         Prediction type                & ground-truth class & predicted class \\ \hline
         True positive for class i  & i              & i           \\
         False positive for class i & j\(\neq\) i              & i  \\
         False negative for class i & i              & j \(\neq\) i  \\
    \end{tabular}
    \caption{Definition of true positive, false positive, and false negative predictions for a data point belonging to ground-truth class i.}
    \label{tab:tpfpfn}
\end{table}

The performance of a model can be visualized using a confusion matrix. A schematic for the confusion matrix is shown in Figure \ref{fig:CM}. For confusion matrix \(C\), element \(C_{i,j}\) shows the number of samples with ground truth class \(i\) that were predicted to belong to class \(j\). Elements on the diagonal represent correct predictions, and all off-diagonal elements are incorrect. For class \(k\), true positive predictions are on element \(C_{k,k}\); false positives are all other elements in column \(k\), \(C_{j \neq k,k}\); and false negatives are all other elements in row \(k\), \(C_{k,j\neq k}\). Thus, the confusion matrix gives a convenient and easy-to-interpret representation of the model performance. In Python, the command \textbf{sklearn.metrics.confusion\_matrix()} can be used to compute the confusion matrix for a set of predictions. The precision and recall can be inferred from the confusion matrix or displayed directly with the function \textbf{sklearn.metrics.classification\_report()}.

\section{Results}
\subsection{Standard Analysis}
The standard analysis consists of the following steps in order:
\begin{enumerate}
    \item Contrast limited adaptive histogram equilization is applied to each image.
    \item Each image is resized to 224x224 pixels.
    \item Each image is passed through VGG16 trained with ImageNet weights, and the outputs of the FC1 layer are saved as feature representations. 
    \item The features are transformed using PCA with 50 whitened components, which preserve 73.6\% of the total variance.
    \item Features are clustered via k-means clustering with 7 clusters, k-means++ initialization, and 500 different initialization steps.  
    \item The clustering with the lowest total inertia (not necessarily the greatest accuracy) is used as the final result.
\end{enumerate}

\begin{figure}
    \centering
    \includegraphics{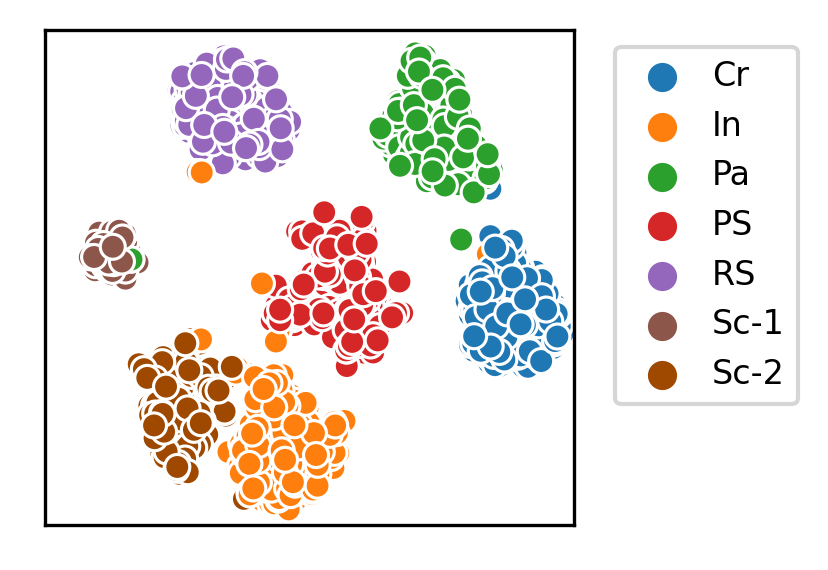}
    \caption{t-SNE projection of feature data colored by cluster identities determined from k-means clustering with 7 clusters from one trial during the standard analysis.}
    \label{fig:std t-SNE}
\end{figure}
This procedure was iterated 10 times with different initial random seeds to verify the repeatability of the approach. The model achieved an average classification accuracy of \(99.40\% \pm 0.16\%\), with minimum accuracy of 99.06\%. The results represent an improvement in both the classification accuracy and variance compared to previous unsupervised methods \cite{Kitahara2018}, and are comparable to supervised methods without the requirement for image labeling. To better interpret the performance, the results for one trial are reported in detail below.

Figure \ref{fig:std t-SNE} shows the t-SNE projection of the feature data colored by the predicted labels determined from one trial during the standard analysis. The color scheme is the same as Figure \ref{fig:tsne_scratch}, with the exception that the extra cluster for scratches is shown in a different shade of brown and denoted `Sc-2.' This allows for easy visual comparison between the ground truth labels and the labels determined from the standard analysis. The labels determined from cluster analysis show good agreement with the ground truth and also the clusters that appear on the t-SNE projection. 

\begin{figure}[hbt]
    \centering
    \includegraphics{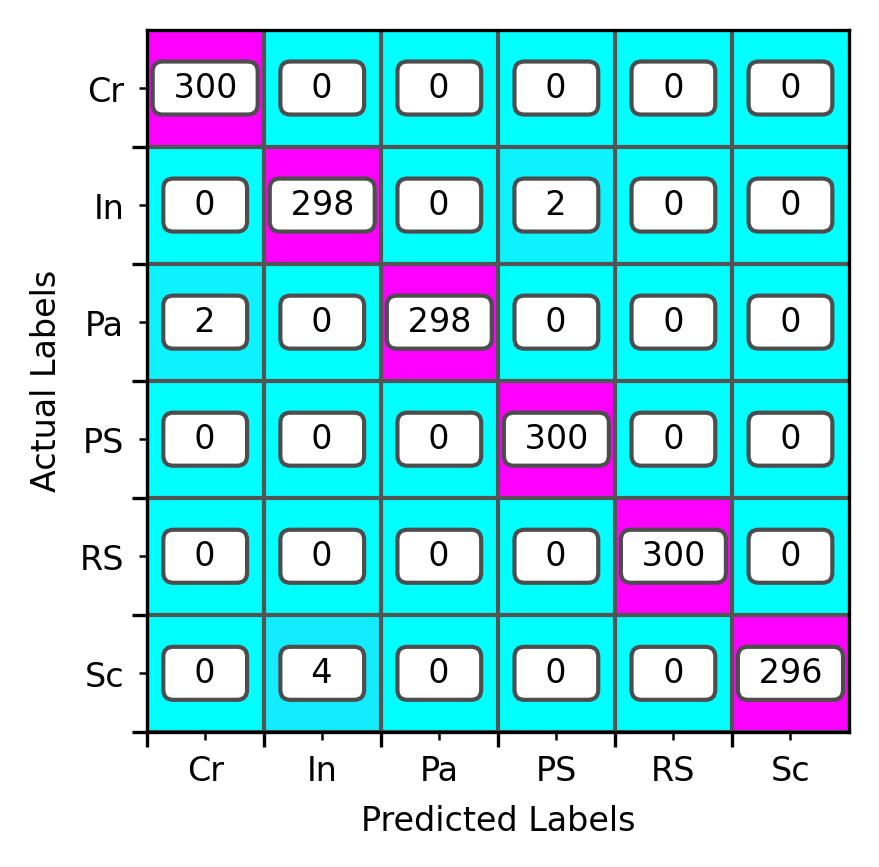}
    \caption{Confusion Matrix for one trial during the standard analysis.}
    \label{fig:std}
\end{figure}

The classification scores from the same trial are visualized in Figure \ref{fig:std} and summarized in Table \ref{tab:std}. This trial achieves 99.6\% classification accuracy for the dataset. The performance can be more precisely understood by looking at the results for each class individually.  The strong diagonal in the confusion matrix indicates good overall classification performance. The model achieves perfect classification of rolled-in scale (RS) images, with no false positives or false negatives. The model also achieves perfect precision for patches (Pa) and scratches (Sc), and perfect recall for crazing (Cr) and pitted surfaces (PS).  The biggest source of confusion is classifying images of scratches to be inclusions, with 4 misclassification errors.  From the t-SNE map in Figure \ref{fig:tsne_scratch}(a), the cluster for vertical scratches is close to the cluster for inclusions, indicating visual similarity between some images of these defect classes. In Figure \ref{fig:SDD}, the bottom two images of inclusions show inclusions that are elongated and oriented vertically and therefore look similar to scratches. The other two sources of errors were predicting patches to be crazing and inclusions to be pitted surfaces, with 2 misclassification errors each.

\begin{table}[htb]
    \centering
    \begin{tabular}{lcc}
    	Class	& Precision	& Recall	\\  \hline 
    	Cr	    & 0.993 	& 1.000 	\\
    	In	    & 0.987 	& 0.993 	\\
    	Pa	    & 1.000 	& 0.993 	\\
    	PS	    & 0.993 	& 1.000 	\\
    	RS	    & 1.000 	& 1.000 	\\
    	Sc	    & 1.000 	& 0.987 	\\
    \end{tabular}
    \caption{Precision and recall for each class for one trial of the standard analysis}
    \label{tab:std}
\end{table}

\subsection{Predictive Model}
\label{SS:predicitive model}
Five-fold cross validation was used to test the performance of the model on data not used to compute the original cluster centers. The image data was randomly divided into 5 equal subsets. Four subsets, denoted the training set, were used to find the PCA components and cluster centers using the standard analysis. The last subset, denoted the validation set, was held out to test the performance of the clusters on data that the model has not seen before. The class labels for each point in the validation set is determined by its closest cluster center in feature space.  This process was repeated 5 times, where each subset was used as the test set once.

The results of the analysis with cross-validation are shown in Table \ref{tab:xval}. On the first set the model achieves a training accuracy of 0.988 and a validation accuracy of 0.983, but on the rest of the trials the training and validation accuracy is at least 0.99.  This indicates that the centroids determined from the training set are reasonable approximations of the true cluster centers for each defect class. Since representations in feature space  can be generated for new images, the model can generalize to classify unseen data with good performance.
Note that this is not possible when clustering points on t-SNE projections, as new points cannot be added without re-computing the whole map. Thus, clustering PCA extends the performance of the method presented in \cite{Kitahara2018} with the ability to classify new images.
\begin{table}[h]
    \centering
    \begin{tabular}{c c c}
         Subset & Train Accuracy & Validation Accuracy  \\ \hline
            1   & 0.988          & 0.983                \\
            2   & 0.990          & 0.997                \\
            3   & 0.992          & 0.994                \\
            4   & 0.992          & 0.994                \\
            5   & 0.993          & 0.992                \\
    \end{tabular}
    \caption{Performance of cluster analysis with 5-fold cross validation.}
    \label{tab:xval}
\end{table}

\subsection{Sensitivy Analysis}

Because the selection of operations and parameters can significantly affect model performance, we conducted a sensitivity analysis for the pre-processing, feature extraction, and clustering elements of the standard analysis. Since there are few known best practices for constructing machine learning systems, this is a recommended step for model optimization. It is important to note that the specific selections made here may not represent the best choices for other image datasets or machine learning architectures, but the methodology of systematically examining the operations and parameters in the analysis pipeline is generally applicable.

\subsubsection{Preprocessing: Histogram Equalization}

\begin{figure}[hbt]
    \centering
    \includegraphics[scale=1]{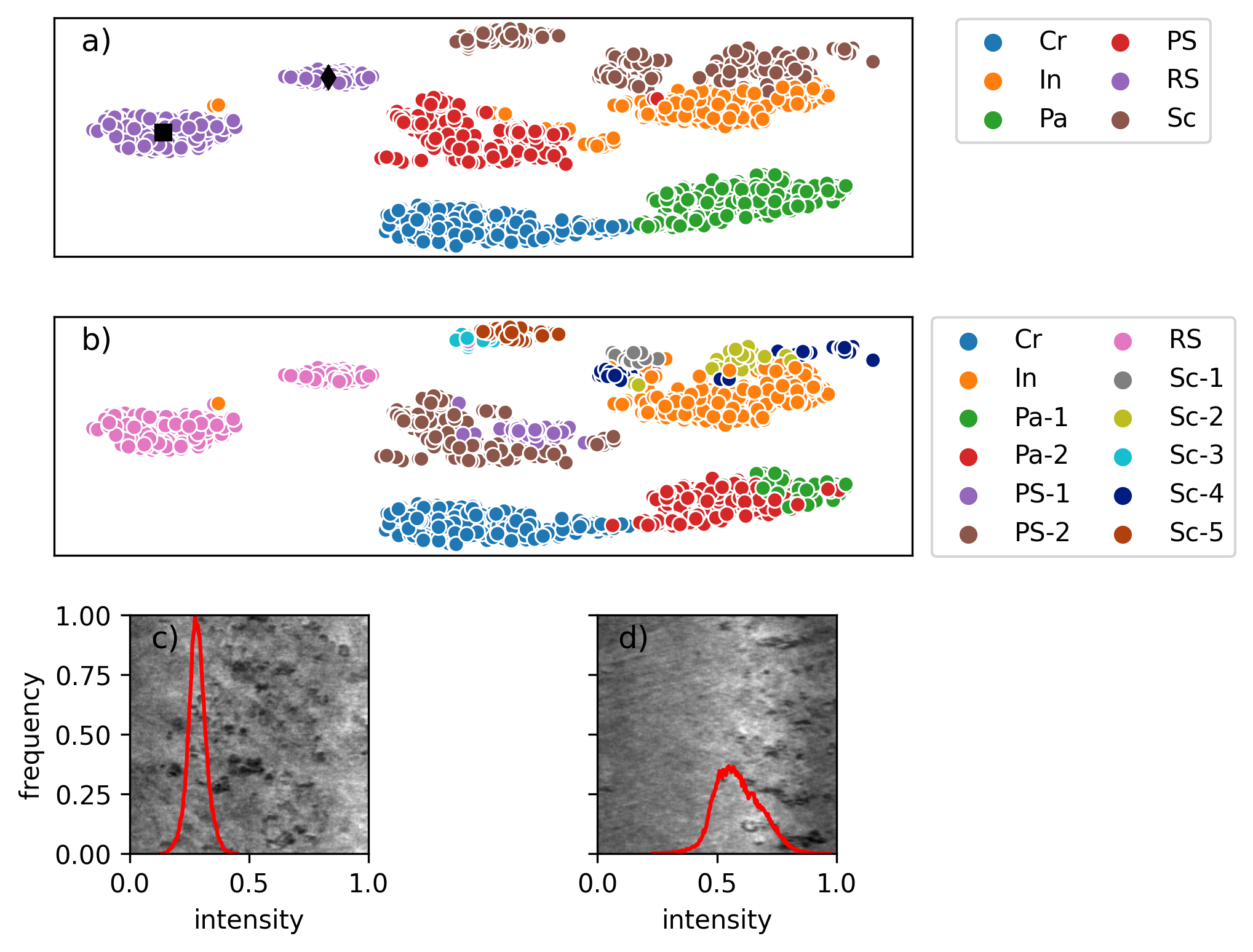}
    \caption{a) t-SNE projection of features without histogram equalization applied. Colors correspond to ground truth labels. b) t-SNE projection of features without histogram equalization applied. Colors correspond to clusters determined from k-means.
    c) Sample image from first rolled-in scale cluster denoted by the black diamond on the t-SNE map. Intensity histogram is overlaid on the image. d) Sample image from second rolled-in scale cluster denoted by the black square on the t-SNE map. The intensity histogram for each image is overlaid on the figure.}
    \label{fig:no ahe tsne}
\end{figure}

\begin{figure}[hbt]
    \centering
    \includegraphics[scale=1]{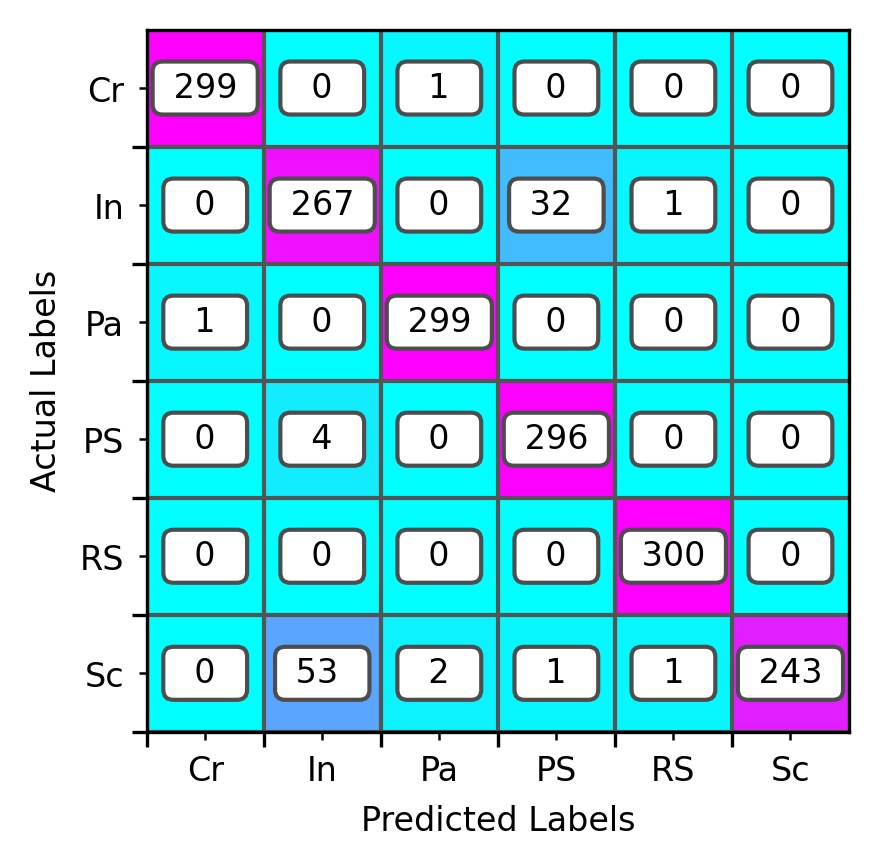}
    \caption{Confusion Matrix for analysis without histogram equalization and clustering with 12 centroids.}
    \label{fig:CM no ahe}
\end{figure}

\begin{table}[htb]
    \centering
    \begin{tabular}{lcc}
    	Class	& Precision	& Recall	\\  \hline 
    	Cr	& 0.997 	& 0.997 	\\
    	In	& 0.824 	& 0.890 	\\
    	Pa	& 0.990 	& 0.997 	\\
    	PS	& 0.900 	& 0.987 	\\
    	RS	& 0.993 	& 1.000 	\\
    	Sc	& 1.000 	& 0.810 	\\
    \end{tabular}
    \caption{Precision and recall for each class with no histogram equalization and clustering with 12 centroids.}
    \label{tab:no ahe}
\end{table}

To determine the impact of histogram equalization on the classification performance, the analysis was repeated on images without applying histogram equalization. The t-SNE projection for the feature representations of these images colored by ground truth labels is shown in Figure \ref{fig:no ahe tsne}(a). Compared to the t-SNE map in the standard analysis, the clusters are more elongated, indicating higher visual variance within each class. Also, the rolled-in scale (RS) cluster has split into 2 distinct clusters. The images in Figures \ref{fig:no ahe tsne}(c) and (d) show typical images from each of the two clusters with their intensity histograms overlaid on the image. The image in Figure \ref{fig:no ahe tsne}(c) has more scale, which appears to be very dark with a less reflective surface. Its intensity distribution is narrow with a maximum value at relative intensity of 0.27. In contrast, the image in Figure \ref{fig:no ahe tsne}(d) has less scale and a more reflective surface. Its intensity histogram is much wider with a peak value at a relative intensity of 0.55. The differences in intensity profiles are captured in the feature representation of the images and appear as two distinct clusters in the t-SNE map.

Applying k-means with 7 clusters results in a classification accuracy of 93\%, which is a significant decrease in performance compared to the approach with histogram equalization. Since the t-SNE map indicates that the data are more spread out, clustering was repeated while varying the number of centroids between 7 and 15. Classification accuracies of these models vary between 87\% and 95\%. The lowest and highest scoring models have 8 and 12 centroids, respectively. The t-SNE map with points colored by their labels determined from clusteing with 12 centroids is shown in Figure \ref{fig:no ahe tsne}(b). 

Interestingly, despite appearing as two distinct clusters on the t-SNE map, images containing the rolled-in scale defect are grouped into a single cluster in feature space. On the other hand,  5 out of 12 clusters are associated with scratches, indicating that removing histogram equalization significantly increases the variance between these images. From the example images in \ref{fig:SDD}, images of scratches have very dark backgrounds, and the scratches themselves are very bright. Thus, the intensity profile heavily depends on the size and number of scratches in the image. Without the use of histogram equalization, these images generate very different visual signals from each other when analyzed with the VGG16 network. 

The classification performance without histogram equalization is shown in Figure \ref{fig:CM no ahe} and summarized in Table \ref{tab:no ahe}. Compared to the standard analysis the decrease in performance is driven by predicting scratches to be inclusions (53 misclassifications) and predicting inclusions to be pitted surfaces (32 misclassifications.) Even with 5 cluster centers for scratches, the model has trouble classifying scratches. This suggests that the removal of histogram equalization causes overlap in feature space between the clusters for scratches and inclusions, as well as inclusions and pitted surfaces, resulting in many classification errors. Since the defect class is independent of the intensity profile of each image, applying histogram equalization is necessary for eliminating the effect that relative brightness has on the visual signal captured in each image.

\subsubsection{Feature Extraction: Choice of Output Layer}

\begin{figure}[p]
    \centering
    \includegraphics{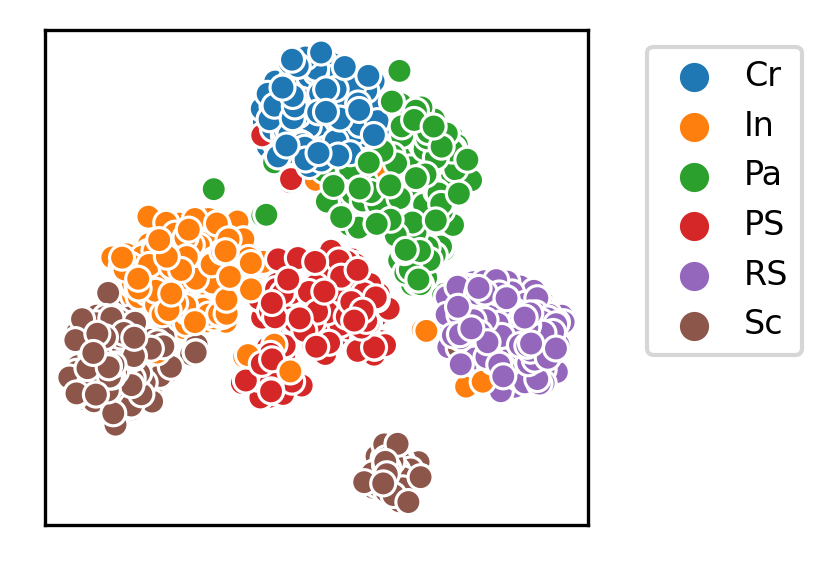}
    \caption{t-SNE projection of feature data extracted from block5\_pool layer from VGG16 with 110 unwhitened PCA components.}
    \label{fig:results b5ptsne}
\end{figure}

    It has been observed that image classification accuracy depends on the choice of the CNN output layer selected as the feature descriptor, and that different image types are best represented by different layers \cite{Ling2017}. Thus, the analysis was repeated using the outputs of the VGG16 fc2 and block5\_pool layers in place of the fc1 layer.  The fc2 fully-connected layer has the same size as the fc1 layer, generating 4096-dimensional features for each image. Simply conducting the standard analysis except replacing the fc1 features with fc2 features results in similar classification performance. The accuracy is 99.4\%, and the biggest source of error is classifying 5 images of scratches to be inclusions. The outputs of the block5\_pool layer are much larger, with features in 25,088 dimensions. Thus, the parameters of the analysis are changed to maximize the classification performance. For the block5\_pool layer, 110 PCA components were kept, preserving about 55\% of the total variance of the data. 
    
\begin{figure}[p]
    \centering
    \includegraphics[scale=1]{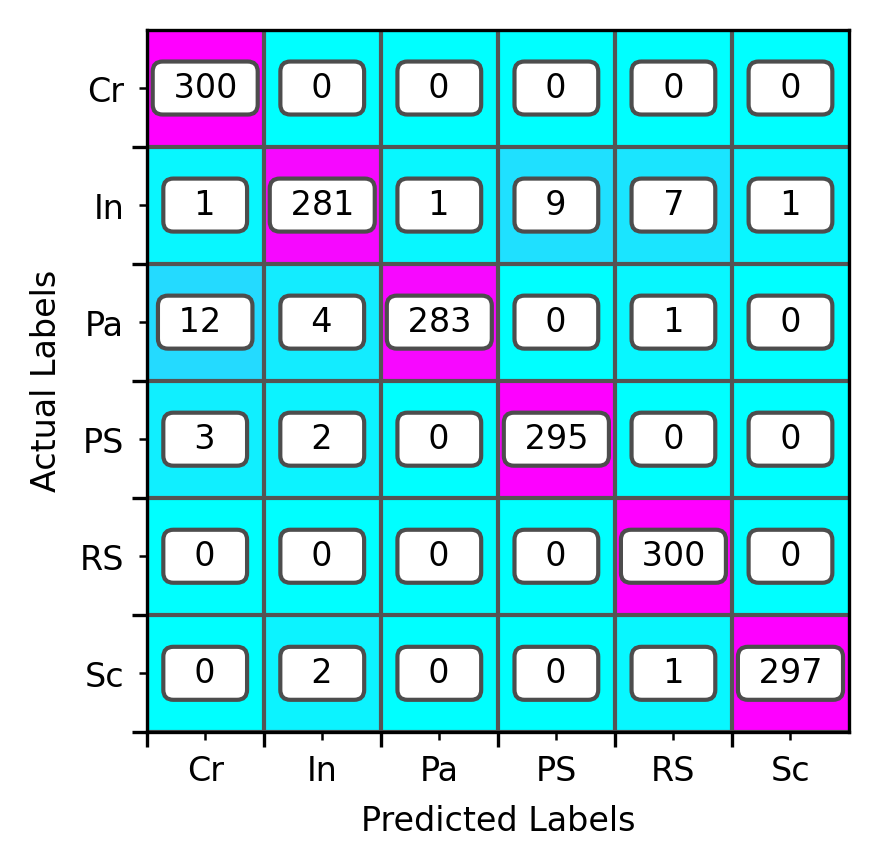}
    \caption{Confusion matrix for clustering block5\_pool features on the t-SNE map.}
    \label{fig:b5p-cm}
\end{figure}

    The t-SNE map of the features colored by their ground truth labels is shown in Figure \ref{fig:results b5ptsne}. The data still cluster by defect type, though some of the clusters appear closer to each other. KMeans was run while varying the number of clusters between 6 and 19. The maximum accuracy was achieved with 9 clusters. However, the accuracy was only 89\%. Interestingly, the decrease in accuracy is driven by 178 predictions of scratches to be inclusions. This decreased the recall of scratches to 0.4 and the precision of inclusions to 0.64. All other classes were correctly labeled with precision and recall values above 0.976. Despite showing good clustering on the t-SNE map, KMeans is unable to resolve scratches from inclusions in feature space. Noting the apparent strong clustering in the t-SNE map, clustering was performed using the t-SNE map directly. With 7 clusters, the model achieves 96.4\% accuracy. Increasing k to 23 improves the cluster accuracy to 97.6\%. The confusion matrix for this is shown in Figure \ref{fig:b5p-cm}. Interestingly, this model only classified two scratches as inclusions, demonstrating improved recall for scratches compared to the k-means analysis. However, the model made more mistakes when classifying images of patches and inclusions, contributing to the lower overall accuracy. The results indicate that the original block5\_pool features contain significant noise, making it difficult for PCA/KMeans to directly capture the clusters. The nonlinear mapping applied during t-SNE helps resolve the signal, especially when separating scratches from inclusions. A neural network can be thought of as an encoder-decoder signal processor. The convolution blocks in VGG16 encode images with lots of filter responses but also have lots of noise. Thus, after extracting features from these blocks, additional techniques such as t-SNE are needed to de-noise the useful signal for cluster analysis. In contrast, the fully connected layers extract signal from the noisy outputs of the convolution blocks, so PCA and k-means can be applied directly to these features without additional transformations.

\subsubsection{Feature Extraction: PCA Whitening and Number of Components}
\label{sec:results-pca}

\begin{figure}[bt]
    \centering
    \includegraphics[scale=1]{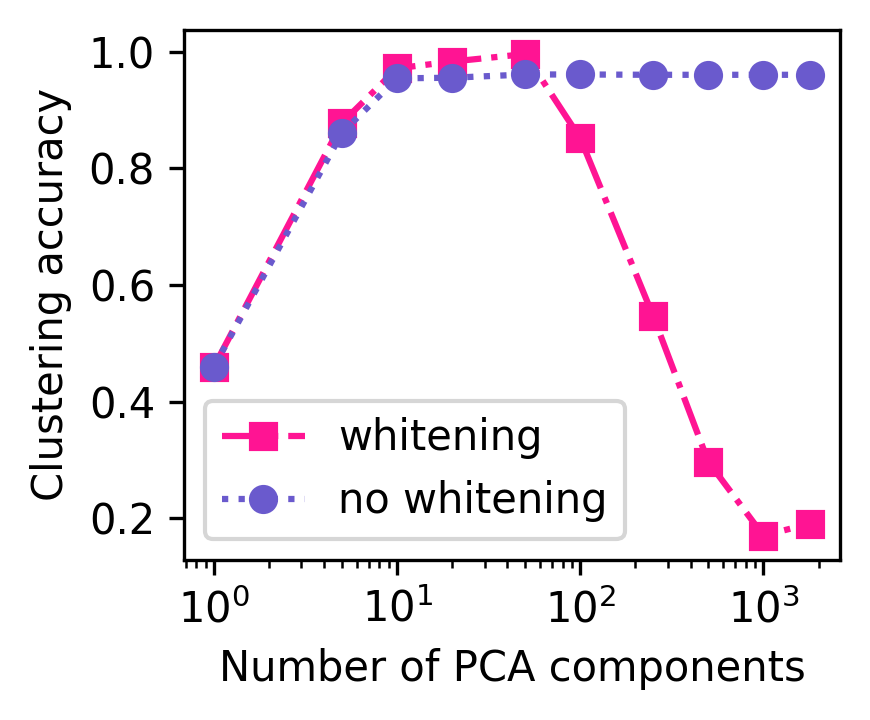}
    \caption{Clustering accuracy vs number of PCA components used for both whitenind and unwhitened components.}
    \label{fig:pca_n}
\end{figure}

The standard analysis uses PCA with 50 components followed by whitening to compute the final feature representation of the images. To determine how classification performance changes with the number of components used, the analysis was conducted while varying the number of components, using both whitened and unwhitened PCA components. Figure \ref{fig:pca_n} shows the results. For small numbers of components, the results with and without whitening are consistent with each other. With only one component, the model still achieves 46\% accuracy. As the number of components increases to 10, the accuracy increases to around 96\%. Without using whitening, the accuracy plateaus at 96\% as the number of components is increased to 1800. After the first 10 components, each additional component contributes a very small amount of variance and therefore does not affect the cluster results. 

The results with whitening demonstrate a different trend. As the number of components is increased to 50, the accuracy increases to a maximum value of 99.6\%.  However, continuing to increase the number of components causes a sharp drop in classification performance. Using 1000 components results in a classification accuracy of 17\%, which is about equal to the expected accuracy for random guessing. Whitening normalizes the variance across all PCA components to unit value. Thus, components that explain less of the variance in the original data but still contain useful signal are able to contribute to clustering the data, resulting in improved classification performance. However, if too many components are included, components that only contain noise drown out the useful signal. With far too many noisy components, the signal is entirely lost. Thus, whitening can significantly improve the classification performance of clustering algorithms like k-means, but can also introduce significant error if too many components are used. 

Although whitening can increase the clustering performance, it also increases the chance that k-means gets stuck in a local minimum and returns sub-optimal classification results. Because of this, including whitening in the analysis increases the number of cluster initialization steps that should be used when running k-means. This is discussed in more detail in Section \ref{ss:cluster}.

\subsubsection{Clustering: Initialization and whitening}
\label{ss:cluster}
\begin{figure}[!htb]
    \centering
    \includegraphics{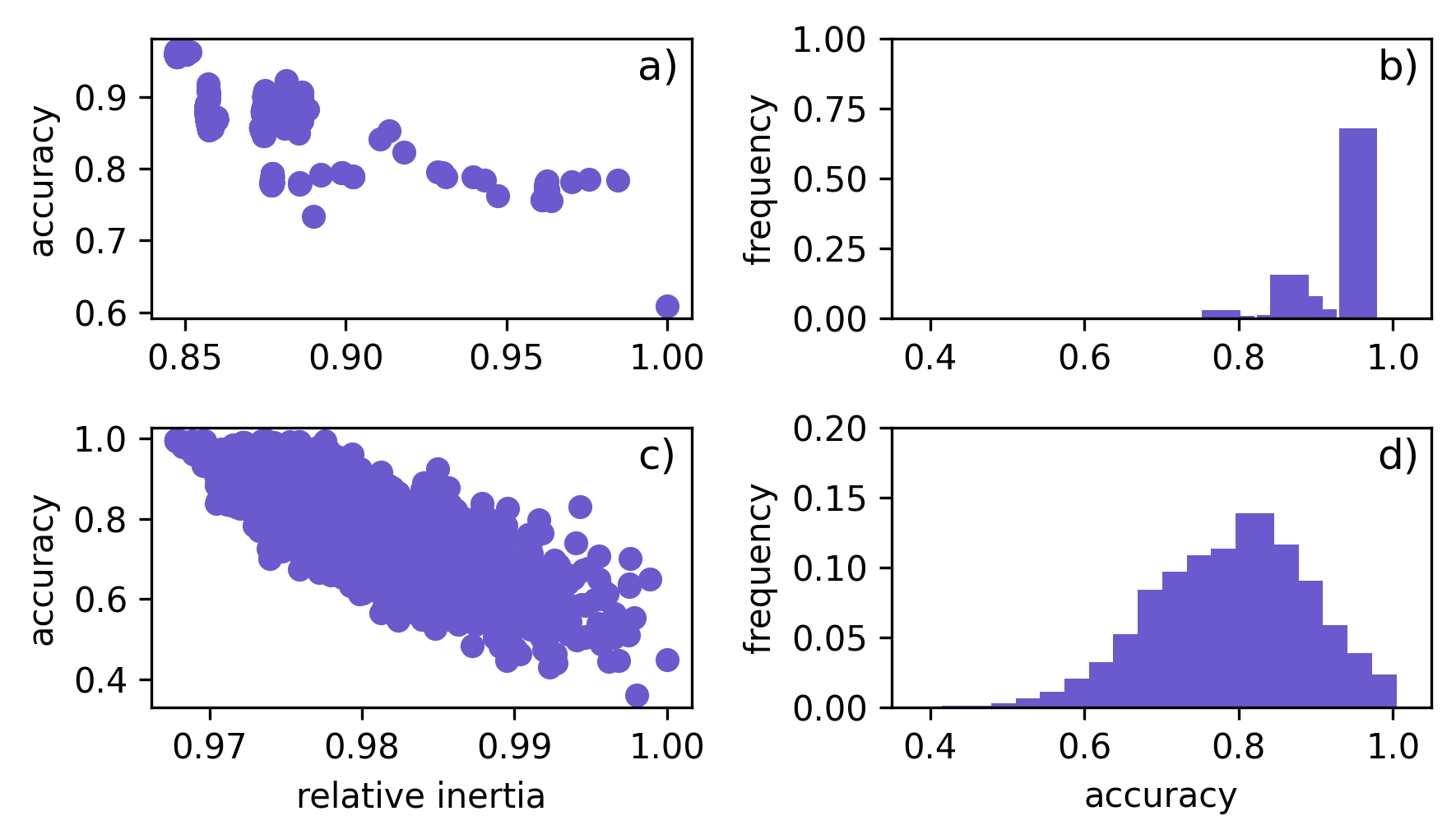}
    \caption{a) Classification accuracy versus relative inertia for 5000 trials of k-means clustering on unwhitened PCA components. b) Histogram of classification accuracy scores for the 5000 trials of k-means clustering on unwhitened PCA components. c) Classification accuracy versus relative inertia for 5000 trials of k-means clustering on whitened PCA components. d) Histogram of classification accuracy scores for the 5000 trials of k-means clustering on whitened PCA components.}
    \label{fig:kmeans_n_iter}
\end{figure}

To determine the impact of the initial choice of cluster centers, k-means was run 5,000 times with different initialization steps, and the final cluster accuracy and inertia were recorded. This analysis was conducted for feature representations with and without whitening applied. Figures \ref{fig:kmeans_n_iter}(a) and (b) show the clustering accuracy versus relative inertia and the histogram of accuracy scores, respectively, for the analysis performed on unwhitened components.  
Depending on the choice of initial cluster centers, the inertia of the final clusters varies by 15\% and the accuracy ranges from 0.61 to 0.96.  The accuracy of the point with the lowest inertia is 0.961, which is close to the maximum accuracy achieved. The correlation between accuracy and inertia is -0.76, indicating that clustering with lower inertia generally results in a higher classification accuracy for this dataset. The scores are irregularly distributed. 68\% of the trials achieved higher than 93\% accuracy. This indicates that there is a strong minimum in inertia, and not that many iterations of k-means are required to achieve a good clustering performance.

Figures \ref{fig:kmeans_n_iter}(c) and (d) show the clustering accuracy versus relative inertia and the histogram of accuracy scores, respectively, for the analysis performed on whitened components. The relative inertia for the tests with whitening is much smaller, spanning only about 3\% between all trials. Despite this, there is a larger range in classification accuracy, ranging from 0.36 to 0.96. The accuracy of the model with the lowest inertia is 0.996. As expected, whitening allows the model to reach higher classification accuracy. Similar to the results for unwhitened components, accuracy has a negative correlation with inertia with a correlation coefficient of -0.79.   Unlike the experiment without whitening, the scores for each trial are more normally distributed, with most trials reaching near 85\% accuracy. Only 0.7\% of trials achieved accuracies higher than 99\%. Thus, when whitening is applied, there are many local minima in inertia in which KMeans can get trapped.  The results indicate that despite increasing the potential for high classification performance, whitening PCA components also increases the variance between different trials of k-means. Thus, in order to have a high likelihood of finding good clustering with whitened PCA components, k-means needs to be run with many initial centroid selections.

\clearpage 

\section{Conclusions}
This paper provides an in-depth description of the steps required to apply transfer learning to classify images in the North Eastern University Steel Surface Defects Database with k-means clustering. The approach outlined in this study achived \(99.4\% \pm 0.16\%\), demonstrating improved accuracy compared to previous studies in the literature.  A sensitivity analysis was conducted to demonstrate the impact of each step in the analysis on the results. Histogram equalization improves classification performance by reducing differences between images with the same defects but different brightness profiles. Using the outputs of the fully connected layers in VGG16 maximizes the signal-to-noise ratio of the feature descriptors, resulting in a useful and relatively compact feature description of each image. Using PCA with enough components to preserve about 75\% of the total variance and applying whitening optimized the classification performance by maximizing the useful signal captured in the feature descriptors. Despite there being only 6 defects, running k-means with 7 clusters was needed to account for both veritical and horizontal scratches.  Clustering whitened PCA components results in the maximum classification performance but also increases the variance in performance of individual trials. Thus, to maximize the classification accuracy, k-means was run with many different initialization steps, and the model with the lowest total inertia was used. Finally, because the analysis does not rely on clustering points on a t-SNE map, the k-means model can be used to classify new images with accuracy above 99\%. This allows for automated classification of large numbers of images in high-throughput experiments and quality control applications.

\section*{Acknowledgments}

This work was supported by the National Science Foundation under grant CMMI-1826218 and by the Air Force Research Laboratory under cooperative agreement number FA8650-19-2-5209. 

\section*{Disclaimer}

The views and conclusions contained herein are those of the authors and should not be interpreted as necessarily representing the official policies or endorsements, either expressed or implied, of the Air Force Research Laboratory or the U.S. Government. 

\printbibliography

\end{document}